\documentclass[amsmath,amssymb,aps,prl,reprint, showpacs, superscriptaddress]{revtex4-1}
\usepackage{graphicx}
\usepackage{amsthm}
\usepackage{amsmath}
\usepackage{amssymb}
\usepackage{dsfont}
\usepackage{bm}

\usepackage{color}
\begin{document} 
\title[]{Maximum efficiency of steady-state heat engines at arbitrary power}
\author{Artem Ryabov} 
\email{rjabov.a@gmail.com} 
\author{Viktor Holubec}
\email{viktor.holubec@mff.cuni.cz}
\affiliation{ 
 Charles University in Prague,  
 Faculty of Mathematics and Physics, 
 Department of Macromolecular Physics, 
 V Hole{\v s}ovi{\v c}k{\' a}ch 2, 
 CZ-180~00~Praha, Czech Republic 
} 
\date{\today} 
%%%%%%%%%%%%%%%%%%%%%%%%%%%%%%%%%%%%%%%%%%%%%%%%%%%%%%%%%%%%%%%%%%%%%%%%%%%%%%%%%%%%%%%%%%%%%%%%%%%%%%%%%%%%%%%%%%
\begin{abstract} 
We discuss the efficiency of a heat engine operating in a nonequilibrium steady state maintained by two heat reservoirs. Within the general framework of linear irreversible thermodynamics we derive a universal upper bound on the efficiency of the engine operating at arbitrary fixed power. Furthermore, we show that a slight decrease of the power below its maximal value can lead to a significant gain in efficiency. The presented analysis yields the exact expression for this gain and the corresponding upper bound.
\end{abstract}

% Mechanics statistical, 05.20.-y
% Irreversible thermodynamics, 05.70.Ln
% Heat engines, 07.20.Pe

\maketitle

The Carnot efficiency $\eta_{C}=1-T_{c}/T_{h}$ \cite{Carnot1978, Clausius1856} provides the upper bound on  efficiency of heat engines working between two reservoirs at temperatures $T_h $ and $T_c$, $T_h>T_c$. Though crucial from the theoretical point of view \cite{Callen2006}, practical applications of $\eta_{C}$ are rather limited, since the Carnot efficiency can be reached only when the heat engine operates reversibly. Reversible operation implies extremely long duration of the working cycle. As a result, when the engine efficiency reaches the upper bound $\eta_{C}$, the output power is zero. 
Appealing universality of the upper bound $\eta_{C}$, which depends solely on the two temperatures, and the needs of engineering solutions stimulated an intensive search for a more practical upper bound on the efficiency of heat engines operating \emph{at finite power}. 
A promising candidate for which at least some universal properties can be derived was introduced about half century ago \cite{Yvon1955, Chambadal1957, Novikov1958}, it is the efficiency \emph{at maximum power} $\eta^{\star}$.

The upper bound on the efficiency at maximum power (EMP) in the linear response regime (linear in $\eta_{C}$) is equal to the famous Curzon-Ahlborn \cite{Curzon1975} formula $\eta_{CA}=1-\sqrt{T_{c}/T_{h}}$, which is  to the linear order in $\eta_{C}$ equal to the half of the Carnot efficiency, $\eta_{CA} =\eta_{C}/2 + O(\eta_{C}^{2})$ \cite{BroeckPRL2005}. 
The upper bound $\eta^{\star}=\eta_C /2$ is achieved by a particular class of heat engines with  strongly coupled thermodynamic fluxes. The assumption of \emph{strong coupling} (see discussion below Eq.~\eqref{q}) means that the heat flux is proportional to the flux, which generates work on the surrounding \cite{EspositoPRB2009, EspositoPRL2009, BroeckEurPhysJ2015}.  

In the present study we stay in the linear response regime (linear in $\eta_C$), however, we go beyond the regime of maximum power and study the engine efficiency \emph{at an arbitrary power} $P$, $0\leq P\leq P^{\star}$ ($P^{\star}$ stands for the maximum power). One of the main messages is that the universal bounds on efficiency can be derived for an arbitrary $P$ and not only at the point of maximum power which was considered in several recent studies \cite{BroeckPRL2005, EspositoPRB2009, EspositoPRL2009, BroeckEurPhysJ2015, Schmiedl2008, Esposito2010b, Tomas2013}, see however \cite{Hernandez2007, HernandezPRE2008, Hernandez2010, GuoJApplPhys2012, deTomasPRE2013, LongPRE2012, HolubecRyabov2015, ViktorJSTAT} for optimal regimes other than that with maximum power. 
To this end we introduce relative deviations from the regime of maximum power, the relative gain in efficiency $\delta \eta$ and power~$\delta P$:
\begin{equation} 
\label{delty}
\delta P = \frac{P-P^{\star}}{P^{\star}}, 
\qquad \delta \eta = \frac{\eta-\eta^{\star}}{\eta^{\star}}, 
\end{equation} 
where $ -1 \leq  \delta P \leq 0$. Such normalization of the two principal engine characteristics allows us to derive several explicit results. One of them is that it is possible to provide a universal upper bound for the efficiency at an \emph{arbitrary power} $P$. The bound depends explicitly on $\delta P$ and it reads
\begin{equation} 
\label{upperbound} 
\eta(P) = \frac{\eta_C}{2}  \left( 1 + \sqrt{-\delta P} \right) . 
\end{equation}
At the maximum power regime $\delta P=0$, the above formula reduces to the well known upper bound $\eta_{C}/2$ for the EMP in the linear response theory ($\eta_{C}$ small). On the other hand, for a zero power, i.e., for 
$\delta P \to -1$, Eq.~\eqref{upperbound} yields the Carnot efficiency. 

The upper bound \eqref{upperbound} on the efficiency at arbitrary power paves the way for better understanding of behavior of real-world engines and thermal plants. These devices in most cases do not operate in the regime of maximum power. Instead, the compromise between power and efficiency is chosen since decreasing the power ($\delta P<0$) can significantly enhance the efficiency as compared to $\eta^{\star}$ \cite{Chen2001, DeVos1992, Chen1994, HolubecRyabov2015, Dechant2016}. The upper bound \eqref{upperbound} predicts that significant enhancement can be achieved by a slight decrease of the power, since the relative gain in efficiency as compared to the relative power loss   
\begin{equation}
\label{dEtadP}
\frac{\delta \eta }{(-\delta P)} =  \frac{ 1}{ \sqrt{-\delta P}},  
\end{equation}
diverges for powers near the maximum power, $P \approx P^{\star}$. Again, Eq.~\eqref{dEtadP} represents the upper bound for the relative enhancement of efficiency achieved by strong-coupling models. At an arbitrary coupling the result will differ by a constant model-dependent prefactor, see Eq.~\eqref{Gainq}. Quite remarkably, this significant enhancement of efficiency $\delta \eta \sim \sqrt{-\delta P}$ is observed in several particular models even beyond the linear response regime 
, e.g.~in recent studies on quantum thermoelectric devices \cite{Whitney2014, Whitney2015}, for a stochastic heat engine based on the underdamped particle diffusing in a parabolic potential \cite{Dechant2016} and also for the so called low-dissipation heat engines \cite{HolubecRyabov2016}.

\begin{figure}
	\centering
		\includegraphics[width=0.7\columnwidth]{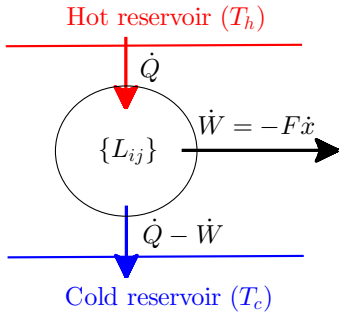}
	\caption{\label{fig:illustr} Steady-state heat engine. Part of the heat flow $\dot{Q}$ from the hot reservoir is transformed by off-diagonal elements of linear relations \eqref{LinResp} into the engine output power $\dot{W}$.} 
\end{figure}

%%%%%%%%%%%%%%%%%%%%%%%%%%%%%%%%%%%%%%%%%%%%%%%%%%%%%%%%%%%%%%%%%%%%%%%%%%%%%%%%%%%%%%%%%%%%%%%%%%%%%%%%%%%%%%%%%%%%%%%%%%%%%%%%%
%%%%%%%%%%%%%%%%%%%%%%%%%%%%%%%%%%%%%%%%%%%%%%%%%%%%%%%%%%%%%%%%%%%%%%%%%%%%%%%%%%%%%%%%%%%%%%%%%%%%%%%%%%%%%%%%%%%%%%%%%%%%%%%%%
{\it Steady-state heat engine.}
We consider the simplest steady-state model of work extraction from the heat flow \cite{BroeckPRL2005}. The model is illustrated in Fig.~\ref{fig:illustr}, it comprises just two thermodynamic forces $X_1$, $X_2$ and fluxes $J_1$, $J_2$. 
The first thermodynamic force $X_1=F/T$ determines the work $W = -F x$ performed by the system on the surrounding, where $x$ is the conjugate variable to $F$ and $T$ stands for the system temperature. In general, the force $F$ can be of mechanical, chemical, or electrostatic origin.
The corresponding thermodynamic flux is $J_1 = \dot{x}$, where the dot denotes the time derivative. The system performs work against $F$ due to the heat flux $J_2=\dot{Q}$ through the system from the hot reservoir to the cold one. The temperature of the hot reservoir $T_{h}$ is larger but comparable to the temperature of the cold reservoir $T_{\rm c}$. The temperature difference $\Delta T = T_{h}-T_{c}$ is assumed to be small as compared to  $T \approx T_{c} \approx T_{h}$, hence we can write the second thermodynamic force $X_2$ to the first order in the relative temperature difference as $X_2 = 1/T_{c}-1/T_{ h} \approx \Delta T/T^{2}$.

%%%%%%%%%%%%%%%%%%%%%%%%%%%%%%%%%%%%%%%%%%%%%%%%%%%%%%%%%%%%%%%%%%%%%%%%%%%%%%%%%%%%%%%%%%%%%%%%%%%%%%%%%%%%%%%%%%
%%%%%%%%%%%%%%%%%%%%%%%%%%%%%%%%%%%%%%%%%%%%%%%%%%%%%%%%%%%%%%%%%%%%%%%%%%%%%%%%%%%%%%%%%%%%%%%%%%%%%%%%%%%%%%%%%%

Within the framework of linear irreversible thermodynamics the forces $X_1$, $X_2$ and fluxes $J_1$, $J_2$ are coupled by the linear relations 
\begin{equation}
\label{LinResp}
J_1 = L_{11}X_{1} + L_{12}X_{2},
\quad
J_2 = L_{21}X_{1} + L_{22}X_{2}.
\end{equation}
Following Ref.~\cite{KedemCaplan1965}, we introduce the ``degree of coupling'' $q$ between the fluxes $J_{1}$ and $J_{2}$, which is defined in terms of the coefficients $L_{ij}$ as 
\begin{equation}
q^{2} = \frac{L_{12}^{2}}{L_{11}L_{22}}, \qquad 
-1 \leq q \leq 1.
\label{q} 
\end{equation}
Physically reasonable values of $q$ follow from the requirement that the entropy production is non-negative, $\dot{S}=J_1 X_1 + J_2 X_2 \geq 0$. This implies for the Onsager coefficients $L_{ij}$ that 
$ 
L_{11} \geq 0$,
$L_{22} \geq 0$,
$L_{11}L_{22}-L_{12}L_{21} \geq 0$ and hence we must have $-1 \leq q \leq 1$. In the case of the so called strong coupling, $q^{2}=1$, the two fluxes are proportional to each other \cite{KedemCaplan1965, BroeckPRL2005, BroeckAdvChemPhys2007, BroeckEurPhysJ2015}. Recently, the idea of strong coupling has been extended beyond the linear-response regime. It is essential for deriving the universal properties of EMP \cite{EspositoPRB2009, EspositoPRL2009, BroeckEurPhysJ2015} for nano-scale heat engines.

The efficiency $\eta$ and the power output $P=-F\dot{x}$ of the engine are defined as 
\begin{equation}
P=\dot{W}=-J_{1}X_{1}T,\qquad
\eta = \frac{P}{\dot{Q}}=-\frac{\Delta T}{T}\frac{J_1 X_1}{J_2 X_2}.
\end{equation}
The efficiency at maximum power in the present model was derived in Ref.~\cite{BroeckPRL2005}. For a fixed temperature difference $\Delta T$ the model contains just one optimization parameter, the external load $F$. The optimal value $X_{1}^{\star}=F^{\star}/T$ of the force $X_{1}$ follows  immediately from the expression for the output power
$P = -(L_{1 1} X_1 + L_{1 2}X_2 ) X_1 T$, which exhibits a maximum for 
\begin{equation}
X_1^{\star}= - \frac{ L_{12}X_{2}} {2 L_{11}}. 
\end{equation}
Thus the maximum power is achieved at the half of the force for which the engine stops (half of the maximal load). The maximum power and the corresponding efficiency are
\begin{equation}
\label{PEtaMP}
P^{\star}= \frac{\eta_{C}^{2}}{4} L_{2 2}  q^{2} T, \qquad \eta^{\star} = \frac{ \eta_{C}}{2} \frac{q^{2}}{2-q^{2}}. 
\end{equation}

%%%%%%%%%%%%%%%%%%%%%%%%%%%%%%%%%%%%%%%%%%%%%%%%%%%%%%%%%%%%%%%%%%%%%%%%%%%%%%%%%%%%%%%%%%%%%%%%%%%%%%%%%%%%%%%%%%
\begin{figure}
	\centering
		\includegraphics[width=0.85\columnwidth]{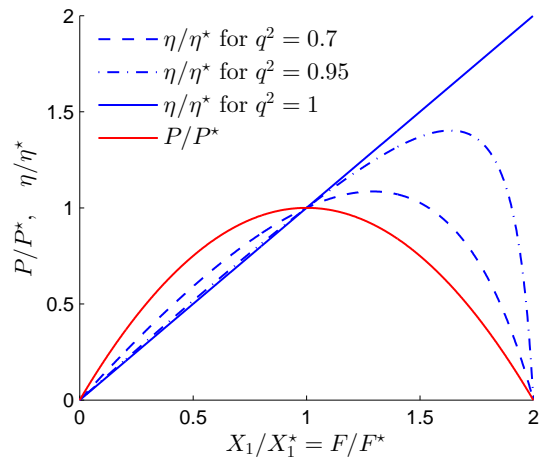}
	\caption{\label{fig:power} Efficiency higher than EMP $\eta^{\star}$ are achieved for higher loads $F>F^{\star}$ (or $X_{1}>X_{1}^{\star}$). On the other hand, when the external force is decreased ($X_{1}<X_{1}^{\star}$), the engine efficiency drops below $\eta^{\star}$. Plotted according to Eqs.~\eqref{PEtaP}.} 
\end{figure}
%%%%%%%%%%%%%%%%%%%%%%%%%%%%%%%%%%%%%%%%%%%%%%%%%%%%%%%%%%%%%%%%%%%%%%%%%%%%%%%%%%%%%%%%%%%%%%%%%%%%%%%%%%%%%%%%%%

It is remarkable that the EMP $\eta^{\star}$ depends on the coupling parameter $q$ only (i.e., the linear coefficients $L_{ij}$ enter the result only in the combination given by $q$). Surprisingly, the efficiency $\eta$ for \emph{any} power $P$, or more precisely for any $\delta P$, can be also given as a function of $q$ only. To see this it is convenient to work with  quantities relative to the point of maximum power \cite{HolubecRyabov2015}.
Then after some algebra we obtain
\begin{equation}
\label{PEtaP}
\frac{P}{P^{\star} }= \left(2- \frac{ X_1}{ X_{1}^{\star}}  \right)\frac{ X_1}{ X_{1}^{\star}}, 
\qquad \frac{\eta}{\eta^{\star}} = \frac{P}{P^{\star} } \frac{2-q^{2}}{2-q^{2} \frac{X_1}{X_{1}^{\star}}}.
\end{equation} 
The relative power $P/P^{\star}$ is given by a simple parabolic relation and it  does not depend explicitly on other model parameters. On the other hand, the normalized efficiency $\eta/\eta^{\star}$ depends, similarly as the efficiency at maximum power \eqref{PEtaMP}, on the coupling strength $q$. 

The two engine characteristics (\ref{PEtaP}) are illustrated in Fig.~\ref{fig:power} for different couplings $q$. Notice that efficiencies higher than EMP $\eta^{\star}$ are achieved for higher external loads  $X_{1}>X_{1}^{\star}$ (or $F>F^{\star}$). On the other hand, when the external force $F$ is decreased below its value $F^{\star}$ (or $X_{1}<X_{1}^{\star}$), the engine efficiency drops below $\eta^{\star}$. 
In order to express the efficiency as a function of the power we first find the relative force to be given by  
\begin{equation}
\frac{X_1}{X_{1}^{\star}} = 1\pm \sqrt{-\delta P}.
\end{equation}
The plus sign corresponds to the favorable case when the external load is increased  and the enhancement of efficiency occurs ($\eta>\eta^{\star}$, $X_{1}>X_{1}^{\star}$). The minus sign describes the opposite branch, where decreasing the power from its maximal value $P^{\star}$ reduces the engine efficiency ($\eta < \eta^{\star}$, $X_{1}< X_{1}^{\star}$).
Using the definition of $\delta P$ from Eq.~\eqref{delty}, we obtain the engine efficiency as the function of the relative power loss $(-\delta P)$:
\begin{equation}
\label{EtaAnyQ}
\frac{\eta}{\eta^{\star}} = (1+\delta P)\frac{2-q^{2}}{2-q^{2}(1\pm \sqrt{-\delta P}  )},
\end{equation}
where again the plus sign corresponds to the region of enhanced efficiency $\eta>\eta^{\star}$, $X_{1}>X_{1}^{\star}$.

In the strong coupling limit we obtain bounds on possible values of the engine efficiency
\begin{equation}
\label{EtaPSC}
\eta(P)  = \frac{\eta_{C}}{2}\left( 1 \pm \sqrt{-\delta P} \right) , \quad q^{2}=1.
\end{equation}
In other words, the lower and the upper bounds on the efficiency at any finite power are simply    
\begin{equation}
\label{EtaPSC2}
\frac{\eta_{C}}{2}  \left( 1 - \sqrt{-\delta P} \right) \leq 
\eta \leq \frac{\eta_{C}}{2}   \left( 1 + \sqrt{-\delta P} \right) .
\end{equation}
At this point we should make a remark concerning simple formulas \eqref{EtaPSC}, \eqref{EtaPSC2}. After the pioneering work \cite{BroeckPRL2005} several linear models were studied with the main focus on the universality of the EMP \cite{Gomez-Marin2006, Wang2012, Apertet2012, Sheng2013, Izumida2014, IzumidaNJP2015}; on the maximum efficiency \cite{Hernandez2007, HernandezPRE2008, Benenti2011, Benenti2013, Brander2013, Jiang2014} or/and on other specific performance characteristics \cite{HernandezPRE2006, HernandezPRE2008, ApertetEPL2012, BizarroPRE2008, BizarroAJP2012, BizarroPRE2012, Wang2016} for specific models. For minimal non-linear irreversible models we refer to Refs.~\cite{IzumidaEPL2012, IzumidaEPL2013, IzumidaPRE2015, WangPRE2016, Ponmurugan2016}, for periodically driven engines see Refs.~\citep{BauerPRE2016, Proesmans2016, Brandner2016}. However, to the best of our knowledge,  the explicit form of maximum (and minimum) efficiency \emph{at a given power} (\ref{EtaPSC})  was not discussed in the literature. The formulas \eqref{EtaPSC}, \eqref{EtaPSC2} represent universal bounds on the efficiency of linear irreversible models. They depend just on the upper bound for the EMP, which for all these models is equal to $\eta_{C}/2$.

Another important general feature encoded in the exact formula (\ref{EtaAnyQ}) is that the engine efficiency can increase significantly when the power is changed slightly from its maximal value. Focusing on the branch of the solution for which $\eta > \eta^{\star}$, we obtain for the relative gain in efficiency: 
\begin{equation} 
\label{Gainq}
\frac{\delta \eta }{(-\delta P)} \approx  \frac{q^{2}}{2 -q^{2} } \frac{1}{\sqrt{-\delta P} }, \qquad \delta P \to 0^{-}.
\end{equation}
The relative gain \eqref{Gainq} diverges when power is close to $P^{\star}$, which means that the gain in efficiency when working near maximum power is much larger then the power loss. The upper bound for this gain, $1/\sqrt{-\delta P}$, is obtained in the strong coupling limit $q^{2}\to 1$.

%%%%%%%%%%%%%%%%%%%%%%%%%%%%%%%%%%%%%%%%%%%%%%%%%%%%%%%%%%%%%%%%%%%%%%%%%%%%%%%%%%%%%%%%%%%%%%%%%%%%%%%%%%%%%%%%%%%%%%%%%%%%%%%%%
%%%%%%%%%%%%%%%%%%%%%%%%%%%%%%%%%%%%%%%%%%%%%%%%%%%%%%%%%%%%%%%%%%%%%%%%%%%%%%%%%%%%%%%%%%%%%%%%%%%%%%%%%%%%%%%%%%%%%%%%%%%%%%%%%
{\it Concluding remarks.}
Universality of the efficiency at maximum power has been discussed rather intensively in recent years. Within the framework of linear irreversible thermodynamics, EMP is bounded by $\eta_{C}/2$. Our present work extends this universal upper bound to engines operating at arbitrary fixed power. The result is given in Eq.~\eqref{upperbound}. 
In future studies it would be interesting to extend the present ideas beyond the linear regime in $\eta_C$.  In the case of EMP, the quadratic term in $\eta_{C}$ turns out to be universal under assumptions of certain symmetries of non-linear response coefficients \cite{EspositoPRB2009, EspositoPRL2009, BroeckEurPhysJ2015}. We believe that similar logic can lead to the universal generalization of our result \eqref{upperbound} beyond the linear regime.  

Equation~(\ref{Gainq}) tells us how much favorable it is to operate the engine at a slightly lower power than at the maximal one. In such a regime, the engine attains considerably larger efficiency than the EMP. The upper bound (\ref{dEtadP})  for the gain in efficiency is obtained in the strong-coupling limit $q^{2}\to 1$, where the equality $\delta \eta = \sqrt{-\delta P}$ holds. For a finite coupling strength, the gain is controlled by the $q$-dependent prefactor. The two results (the upper bound and the actual value of the prefactor)  were derived systematically from the exact expression for the efficiency \eqref{EtaAnyQ}. However, the scaling relation $\delta \eta \sim \sqrt{-\delta P}$ should be valid for a large class of models. To see this, let us consider a small deviation $\varepsilon$ from the point of the maximum power (in the present model $\varepsilon= X_{1}-X_{1}^{\star}$). Since the power attains its maximum at $\varepsilon=0$, the series expansion of the difference $(P-P^{\star})$ starts by the quadratic term, $(P  - P^{\star}) \approx - |c| \varepsilon^{2}$. When the efficiency can be expanded as $(\eta -  \eta^{\star} ) \approx a \varepsilon $, we always have $\delta \eta \sim \sqrt{-\delta P}$. Indeed, such scaling is observed in different unrelated settings \cite{Whitney2014, Whitney2015, Dechant2016, HolubecRyabov2016}. It would be interesting to find an engine for which this scaling is violated. Then one may obtain even stronger gain in efficiency for a slight decrease of power below $P^{\star}$.

Finally, it should be noted that another class of universal results for EMP is known for the so called low-dissipation heat engines \cite{Sekimoto1997, Schmiedl2008, Esposito2010b, Tomas2013}. In our subsequent work  \cite{HolubecRyabov2016} we generalize the present considerations to these systems further clarifying the universality of both the derived bound \eqref{upperbound} and the relation (\ref{Gainq}).

%%%%%%%%%%%%%%%%%%%%%%%%%%%%%%%%%%%%%%%%%%%%%%%%%%%%%%%%%%%%%%%%%%%%%%%%%%%%%%%%%%%%%%%%%%%%%%%%%%%%%%%%%%%%%%%%%%
%%%%%%%%%%%%%%%%%%%%%%%%%%%%%%%%%%%%%%%%%%%%%%%%%%%%%%%%%%%%%%%%%%%%%%%%%%%%%%%%%%%%%%%%%%%%%%%%%%%%%%%%%%%%%%%%%%
\bibliographystyle{apsrev4-1}	% (uses file "plain.bst")
%\bibliography{references}


%%%%%%%%%%%%%%%%%%%%%%%%%%%%%%%%%%%%%%%%%%%%%%%%%%%%%%%%%%%%

\end{document}